\documentclass[runningheads]{llncs}

\usepackage[T1]{fontenc}
\usepackage{graphicx}
\usepackage{comment}
\usepackage{amsmath,amssymb}
\usepackage{color}
\usepackage{url}
\usepackage{hyperref}
\usepackage[T1]{fontenc}
\usepackage[numbers]{natbib}
\usepackage{graphicx}
\usepackage[misc]{ifsym}
\usepackage{enumitem}
\usepackage{makecell}
\usepackage{multirow}
\usepackage{subcaption}
\usepackage{cleveref}
\usepackage{bbm}
\usepackage{amsfonts}
\usepackage{booktabs}

\newif\ifreview
\reviewfalse

\ifreview
	\usepackage{lineno}

	\linenumbers
\fi

\begin{document}

%%%%%%%%%%%%%%%%%%%%% Add submission id, track, and title. %%%%%%%%%%%%%%%%%%%%%

% TODO: Please insert your submission number here
\def\SubNumber{053}

% TODO: Please uncomment the track this paper will be submitted to, comment all other lines
%\def\GCPRTrack{Main Track}
\def\GCPRTrack{Special Track: Pattern recognition in the life and natural sciences}
%\def\GCPRTrack{Special Track: Photogrammetry and remote sensing}
%\def\GCPRTrack{Young Researcher's Forum}
%\def\GCPRTrack{Fast Review Track}

% TODO: Replace with your title
\title{M(otion)-mode Based Prediction of Ejection Fraction using Echocardiograms}
\titlerunning{M-mode Based Prediction of Ejection Fraction}
% You can use \thanks for acknowledgment as in: 
%\title{Title\thanks{XXX}}
%Do not add any acknowledgment to the draft 
% version that is used for the review process.  

\ifreview
	% ANONYMOUS SUBMISSION FOR REVIEW
	% DO NOT MODIFY these for the draft version that is used for the review process.
	\titlerunning{GCPR 2023 Submission \SubNumber{}. CONFIDENTIAL REVIEW COPY.}
	\authorrunning{GCPR 2023 Submission \SubNumber{}. CONFIDENTIAL REVIEW COPY.}
	\author{GCPR 2023 - \GCPRTrack{}}
	\institute{Paper ID \SubNumber}
\else

	\author{Ece Ozkan* \inst{1}\orcidID{0000-0002-9889-6348} (\Letter) 
    \and Thomas M. Sutter* \inst{2} \orcidID{0000-0001-7503-4473} (\Letter) 
    \and Yurong Hu \inst{3}\orcidID{0009-0008-8997-0543} 
    \and Sebastian Balzer \inst{4} 
    \and Julia E. Vogt \inst{2}\orcidID{0000-0002-6004-7770}}
    \authorrunning{E. Ozkan et al.}
    \institute{Department of Brain and Cognitive Sciences, MIT, USA \and Department of Computer Science, ETH Zurich, Switzerland \and Department of Information Technology and Electrical Engineering, ETH Zurich, Switzerland \and Department of Biosystems Science and Engineering, ETH Zurich, Switzerland\\ \Letter \hspace{0.1cm} \email{eoezkan@mit.edu, suttetho@inf.ethz.ch} \\
  * Shared first authorship.
  }
\fi

\maketitle              % typeset the header of the contribution

\begin{abstract}
Early detection of cardiac dysfunction through routine screening is vital for diagnosing cardiovascular diseases.
An important metric of cardiac function is the left ventricular ejection fraction (EF), where lower EF is associated with cardiomyopathy.
Echocardiography is a popular diagnostic tool in cardiology, with ultrasound being a low-cost, real-time, and non-ionizing technology.
However, human assessment of echocardiograms for calculating EF is time-consuming and expertise-demanding, raising the need for an automated approach.
In this work, we propose using the M(otion)-mode of echocardiograms for estimating the EF and classifying cardiomyopathy.
We generate multiple artificial M-mode images from a single echocardiogram and combine them using off-the-shelf model architectures.
Additionally, we extend contrastive learning (CL) to cardiac imaging to learn meaningful representations from exploiting structures in unlabeled data allowing the model to achieve high accuracy, even with limited annotations.
Our experiments show that the supervised setting converges with only ten modes and is comparable to the baseline method while bypassing its cumbersome training process and being computationally much more efficient. 
Furthermore, CL using M-mode images is helpful for limited data scenarios, such as having labels for only 200 patients, which is common in medical applications.

\keywords{Echocardiography  \and M-mode Ultrasound \and Ejection Fraction \and Computer Assisted Diagnosis (CAD)}
\end{abstract}

\section{Introduction}
\label{sec:intro}
Cardiovascular diseases (CVD) are the leading cause of death worldwide, responsible for nearly one-third of global deaths \cite{CVD_who}. 
Early assessment of cardiac dysfunction through routine screening is essential, as clinical management and behavioral changes can prevent hospitalizations and premature deaths. 
An important metric for assessing cardiac (dys)function is the left ventricular (LV) ejection fraction (EF), which evaluates the ratio between LV end-systolic and end-diastolic volumes \cite{bamira2018,ouyang2020}. 

% Motivation of echocardiography
Echocardiography is the most common and readily available diagnostic tool to assess cardiac function, ultrasound (US) imaging being a low-cost, non-ionizing, and rapid technology.
However, the manual evaluation of echocardiograms is time-consuming, operator-dependent, and expertise-demanding. 
Thus, there is a clear need for an automated method to assist clinicians in estimating EF. 
 
% Motivation of M-mode
M(otion)-mode is a form of US, in which a single scan line is emitted and received at a high frame rate through time to evaluate the dynamics to assess different diseases \cite{saul2015}.
M-mode is often utilized in clinical practice e.\,g.\,in lung ultrasonography \cite{avila2018,singh2018} or echocardiography \cite{devereux1984,gaspar2014,skinner2017,hensel2018}.
Since cardiac function assessment relies on heart dynamics, M-mode images can be an excellent alternative to B(rightness)-mode image- or video-based methods. 
However, little effort is directed toward exploiting M-mode images in an automated manner. 

% Motivation of supervised and self-supervised learning and labelling 
Data collection and annotation are expensive for most applications. 
Therefore, learning from limited labeled data is critical in data-limited problems, such as in healthcare.
To overcome this data bottleneck, self-supervised learning (SSL) methods have been recently proposed to learn meaningful high-level representations from unlabeled data \citep{lecun2021,shurrab2022}. 

\noindent \textbf{Related Work} A few existing works \citep{kulhare2018,mehanian2019} reconstruct M-mode images from B-mode videos to detect pneumothorax using CNNs. 
Furthermore, authors in \citep{tian2021} propose an automatic landmark localization method in M-mode images. 
A more related method using M-mode images in an automated manner to estimate EF is \citep{sarkar2020}, which uses single M-mode images in parasternal long-axis view to measure chamber dimensions for calculating EF. 

For automated EF prediction, some previous works exploit either still-images \citep{madani2018,zhang2018,ghorbani2020} or spatio-temporal convolutions on B(rightness)-mode echocardiography videos \citep{ouyang2020}.
However, still-image-based methods have a high variability \citep{ouyang2019}, and video-based methods rely on a complex pipeline with larger models. 
Furthermore, \citep{Muhtaseb2022} uses vision transformers and CNNs to tackle the problem of estimating the LV EF, and \citep{Lagopoulos2022} uses geometric features of the LV derived from ECG video frames to estimate EF.
The authors in \citep{Tromp2022} evaluate ML-based methods in a multi-cohort setting using different imaging modalities. 
In the SSL setting, \citep{Dai2022} propose a contrastive learning framework for deep image regression, which consists of a feature learning branch via a novel adaptive-margin contrastive loss and a regression prediction branch using echocardiography frames as input.

\noindent \textbf{Our Contribution} We propose to extract images from readily available B-mode echocardiogram videos, each mimicking an M-mode image from a different scan line of the heart. 
We combine the different artificial M-mode images using off-the-shelf model architectures and estimate their EF to diagnose cardiomyopathy in a supervised regime.
Using M-mode images allows the model to naturally observe the motion and sample the heart from different angles while bypassing cumbersome 3D models. 
Secondly, we propose an alternative scheme for predicting EF using generated M-mode images in a self-supervised fashion while extending contrastive learning.
We design a problem-specific contrastive loss for M-mode images to learn representations with structure and patient awareness. 
We evaluate both regimes on the publicly available EchoNet-Dynamic dataset (\cite{ouyang2019}) and demonstrate both models' effectiveness. 

To the best of our knowledge, this is the first work on image-based and temporal information incorporating cardiac function prediction methods to estimate EF. 
Furthermore, our method can easily be applied to other problems where cardiac dynamics play an essential role in the diagnosis. 
To ensure reproducibility, we made the code available: \url{https://github.com/thomassutter/mmodeecho}.

\section{Methods}
\label{sec:methods}
\begin{figure}[t]
    \centering
    \includegraphics[width=0.5\linewidth]{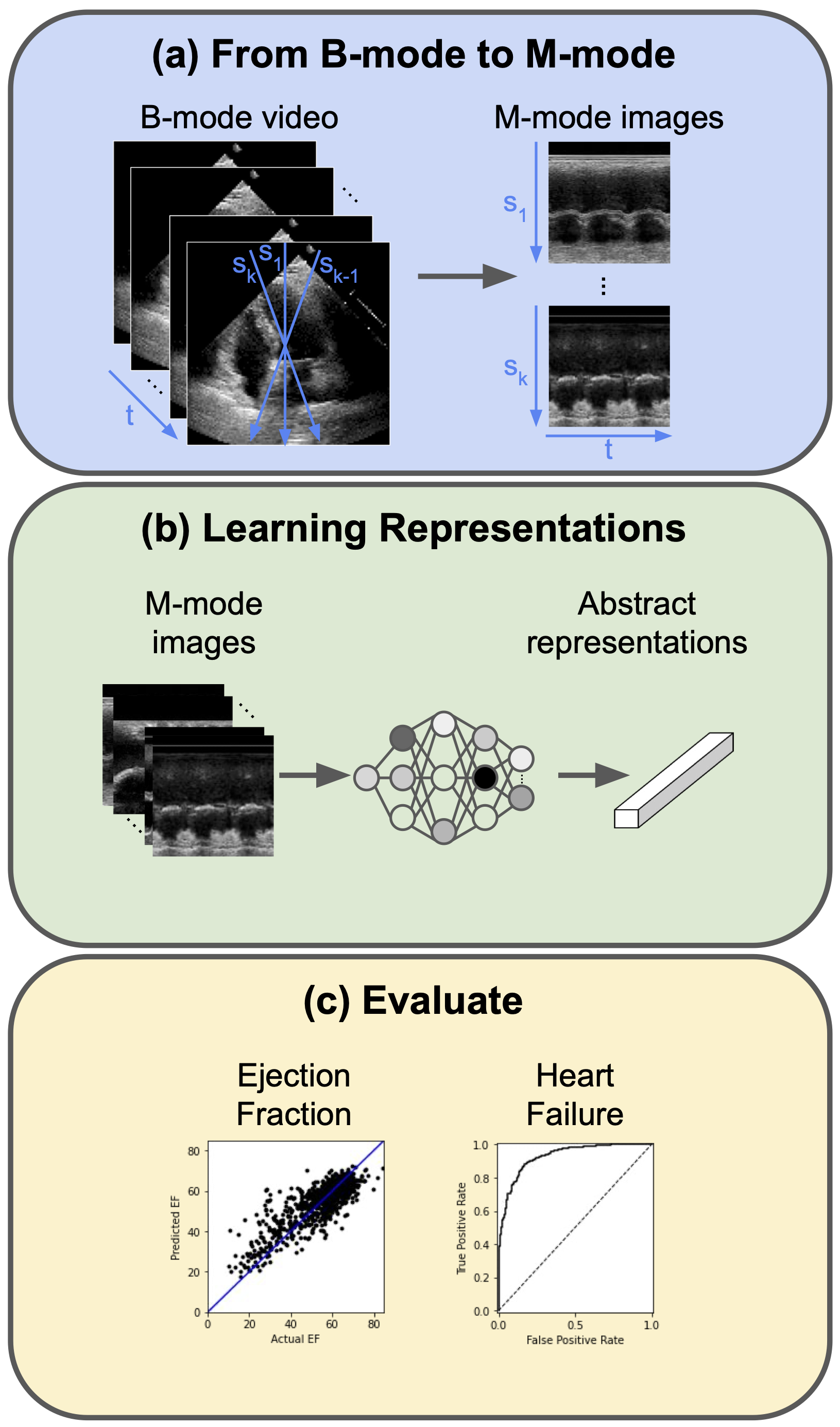}
    \caption{Overview of our proposed method. (a) Generate M-mode images from B-mode echocardiography videos at different scan lines. (b) Learn representations from the generated M-mode images using supervised and self-supervised learning schemes. (c) Evaluate EF prediction to diagnose cardiomyopathy.}
    \label{fig:pipeline}
\end{figure}

This work aims to create a pipeline with as little intervention as possible; thus, our method consists of two parts, as shown in \Cref{fig:pipeline}. 
The first part is extracting M-mode images from readily available B-mode videos. 
The second part includes representation learning, which are lower-level information that preserves more information of the input image and are used to predict EF from M-mode images, including two schemes: supervised and self-supervised learning.  

\subsection{From B-mode Videos to M-mode Images}
\label{subsec:M_mode}

Assume our dataset contains $N$ patients. 
For each patient $i=\{1,2,\cdots,N \}$, the label $y_{i}$ indicates its EF. 
Furthermore, the B-mode echocardiogram video of each patient $i$ is given of size $h \times w \times t$ with $h$ being height, $w$ width, and $t$ number of frames of the video. 
The $m$-th M-mode image of patient $i$ is given as $\boldsymbol{x}_{i}^{m}$ with $m=\{ 1,2,\cdots,M \} $. 
It is a single line of pixels through the center of the image with an angle $\theta_m$ over frames, assuming LV is around the center throughout the video, as in \Cref{fig:pipeline}(a).
This image, corresponding to $\theta_m$, is then of size $s_m \times t$, with $s_m$ as the length of the scan line.
For simplicity, we set $s_m=h \hspace{0.1cm} \forall \hspace{0.1cm} m$ independent of its angle $\theta_m$.
For generating multiple M-mode images, a set of $M$ angles $\boldsymbol{\theta} = [\theta_1, \ldots, \theta_M]$ is used to generate $M$ M-mode images, where the angles $\boldsymbol{\theta}$ are equally spaced between $0^{\circ}$ and $180^{\circ}$.

While the proposed approach for generating M-mode images is intuitive and works well (see \Cref{sec:results_discussion}), other approaches are also feasible.
For instance, the center of rotation in the middle of the image in our M-mode generation process could be changed.
Like that, we could mimic the behavior of the data collection process as every generated M-mode image would resemble a scan line of the US probe.
However, the main goal of this work is to highlight the potential of M-mode images for the analysis of US videos.
Given our convincing results, we leave the exploration of different M-mode generation mechanisms for future work.

\subsection{Learning Representations from M-mode Images}
\label{subsec:learning_representation}    
\subsubsection{Supervised Learning for EF Prediction}
We aim to learn supervised representations using off-the-shelf model architectures to estimate EF. 
Instead of using a single M-mode, one can aggregate the information of M-mode images from the same patient to increase robustness. 
We evaluate two fusion methods for aggregating information among the $M$ M-mode images: early-fusion and late-fusion \cite{baltruvsaitis2018multimodal}.
With early fusion, we construct a $M \times s \times t$ image with the $M$ M-mode images being the $M$ channels of the newly created image.
In late-fusion, we exploit three different methods. For all of the late-fusion schemes, we first infer an abstract representation $\boldsymbol{z}_i^m$ for every M-mode image $\boldsymbol{x}_i^m$.
The representations $\boldsymbol{z}_i^m$ are then aggregated to a joint representation $\boldsymbol{\tilde{z}}_i$ using an LSTM cell \cite{hochreiter1997}, averaging, or concatenating.

We utilize a standard ResNet architecture \cite{he2016deep} with 2D-convolutional layers independent of the fusion principle. 
With 2D-convolutions, we assume a single M-mode image as a 2D gray-scale image with two spatial dimensions, $s$ and $t$.

\subsubsection{Self-Supervised Learning for EF Prediction}
\begin{figure}[t]
    \centering
    \includegraphics[width=0.7\linewidth]{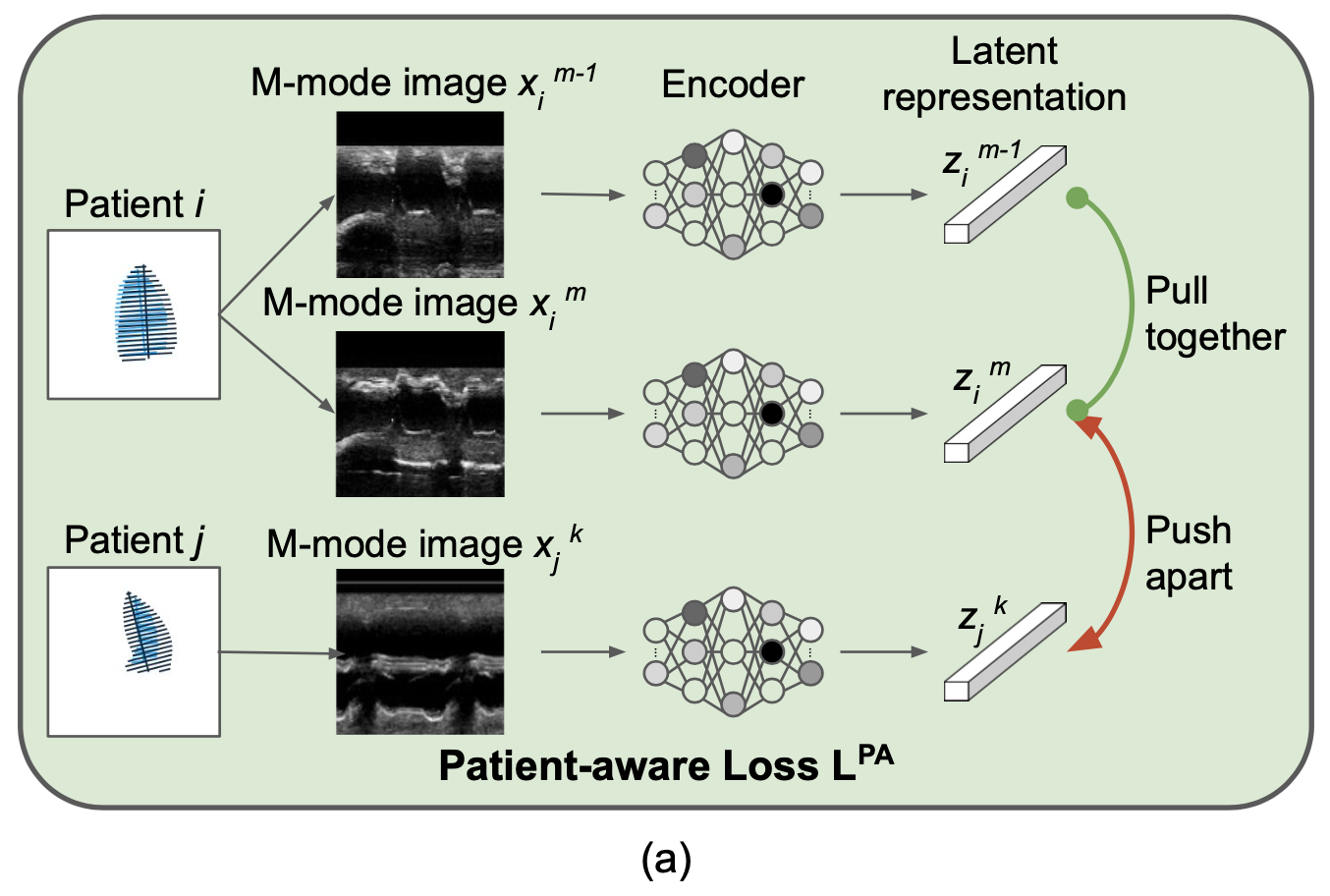} \\
    \includegraphics[width=0.7\linewidth]{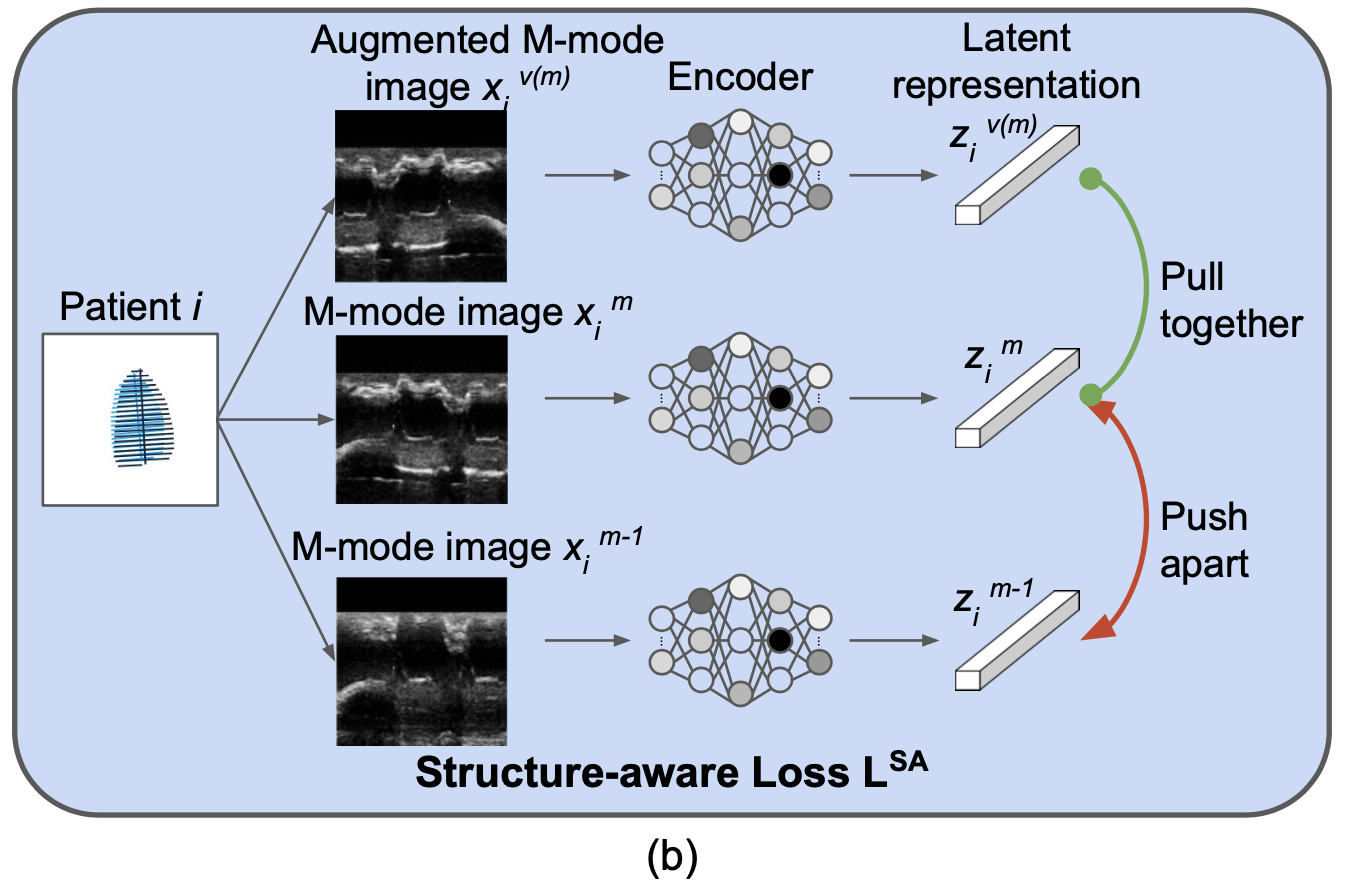}
    \caption{Overview of our proposed SSL method. The contrastive loss includes (a) patient awareness to attract similarity between data from the same patient while discouraging it between different patients and (b) structure awareness to take the (possible) dissimilarity from the same patient into account.}
    \label{fig:SSL_loss}
\end{figure}

This part aims to learn meaningful representations from unlabeled data to estimate EF using echocardiograms.
To this end, we propose an SSL scheme for M-mode images based on contrastive learning, where M-mode images from the same patient can naturally serve as positive pairs since they share labels for many downstream tasks. 
As discussed by \cite{yeche2021}, bio-signal data is inherently highly heterogeneous; thus, when applying learning-based methods to patient data, we need to consider both the similarity and the difference between samples originating from the same patient.
Thus, we propose a problem-specific contrastive loss with patient and structure awareness, as shown in  
\Cref{fig:SSL_loss}.

\begin{figure}[t]
    \centering
    \includegraphics[width=\textwidth]{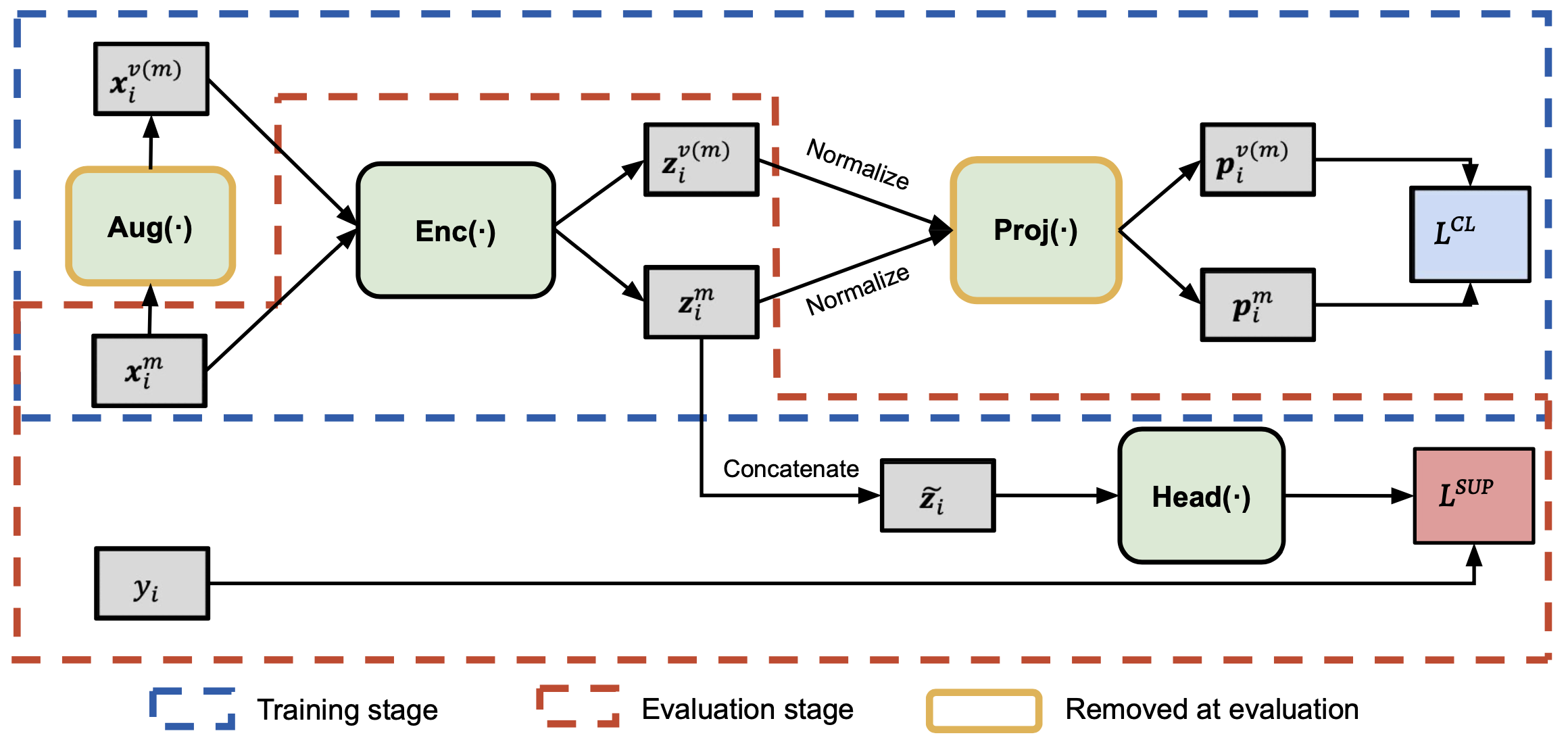}
    \caption{Schema of the contrastive learning framework with training and evaluation stages. The training stage exploits the contrastive loss to learn a representation leveraging the unlabelled images. The evaluation stage exploits these learned representations in a supervised manner to predict EF.}
    \label{fig:SSL_model}
\end{figure}

\paragraph{Contrastive Learning Framework} 
The framework contains training and evaluation stages and the overview is illustrated in \Cref{fig:SSL_model}.
In the training stage, we optimize the model with the contrastive loss leveraging the information from underlying structures of the unlabeled images. 
In the evaluation stage, a multilayer perceptron (MLP) head is trained on top of the learned representations in a supervised manner.

For each generated M-mode image $\boldsymbol{x}_{i}^{m}$, we generate its augmented view $\boldsymbol{x}_{i}^{v(m)}$ using the $Aug(\cdot)$ module.
So the augmented dataset is represented as $\{(\boldsymbol{x}_{i}^{m},\ \boldsymbol{x}_{i}^{v(m)},\ y_{i})\}$. 
The encoder network $Enc(\cdot)$ maps each image $\boldsymbol{x}_i^{m}$ to a feature vector $\boldsymbol{z}_i^m$. 
We utilize a standard ResNet architecture \cite{he2016deep}.

In the training stage, $\boldsymbol{z}_i^m$ is normalized to the unit hyper-sphere before being passed to the projection network. 
Following the work \cite{chen2020}, we introduce a learnable non-linear projection network between the representation and the contrastive loss. 
The projection network $Proj(\cdot)$ takes the normalized lower-level representation $\boldsymbol{z}_i^m$ as input and outputs the higher-level representation $\boldsymbol{p}_i^m$. 
We use a two-layer MLP with ReLU activation as $Proj(\cdot)$ in this work.

In the evaluation stage, we initialize the parameters of the encoder network $Enc(\cdot)$ with the model obtained from contrastive learning and add an MLP head $Head(\cdot)$ to the top. 
For each patient $i$, we have $M$ feature vectors $\boldsymbol{z}_{i}^{m}\in \mathbb{R}^K$. 
The $M$ vectors are then fused to get the joint representation $\boldsymbol{\tilde{z}}_i \in \mathbb{R}^{K \times M}$ and passed to $Head(\cdot)$.
One can have different fusion methods for aggregating information among the $M$ vectors, e.\,g.\,using an LSTM cell \cite{hochreiter1997}, averaging, or concatenating.

\paragraph{Contrastive Loss for M-mode Images} To account for (dis)similarities, we design two loss functions for learning both patient- and structure-awareness. \\

\noindent (a) Patient-aware loss: The goal is to attract the representations from the same patient to be similar while pushing apart representations from different patients (see \Cref{fig:SSL_loss} (a)). 
This enforces two M-mode images to be considered similar if they are from the same patient and dissimilar if they are from different patients. 
The patient-aware loss is given as:

\begin{equation}
    \label{eq:patient_aware_loss}
        L^{PA}=-\frac{1}{M-1}\sum_{i=1}^{N}\sum_{m=1}^{M}\sum_{l\neq m}\log\frac{\exp(\boldsymbol{p}_i^{m}\cdot\boldsymbol{p}_{i}^{l}/\tau)}{\sum_{j,k}\exp(\boldsymbol{p}_i^{m}\cdot\boldsymbol{p}_{j}^{k}/\tau)-\exp(\boldsymbol{p}_i^{m}\cdot\boldsymbol{p}_{i}^{m}/\tau)}
\end{equation}
where $N$ is the number of patients in one batch, $M$ is the number of original M-mode images used for each patient, and $\tau$ is the temperature scaling parameter. 
The term $\boldsymbol{p}_i^{m}$ represents the output of $Proj(\cdot)$.

Inspired by \cite{yeche2021}, we tried defining a neighborhood function % $n(\boldsymbol{x}_{i}^{m}, \boldsymbol{x}_{j}^{l})=(\mathbbm{1}[i=j]\times \mathbbm{1}[|m-l|\leq \lambda])$ 
% and checked if the distance of their mode ordinal numbers are within a certain threshold
to limit the similarity of M-mode images from the same patient. 
However, incorporating neighbourhood to patient-awareness did not further improve the results; 
thus, we used all M-mode images per patient to define the patient-aware loss. \\

\noindent (b) Structure-aware loss: If we only use patient-aware loss $L^{PA}$, there exists a risk that all images from the same patient collapse to a single point \cite{yeche2021}. 
So we propose the structure-aware loss to introduce some diversity (see \Cref{fig:SSL_loss} (b)).
To incorporate this into the learned representations, we construct positive pairs from each M-mode image with its augmentation and consider other combinations as negative pairs. 
It is then defined as:
% \begin{equation}
%     \label{eq:structure_aware_loss}
%     L^{SA}=-\sum_{i=1}^{N}\sum_{m=1}^{2M}\log\frac{\exp(\boldsymbol{p}_i^{m}\cdot\boldsymbol{p}_{i}^{v(m)}/\tau)}{\sum_{l\in N(m)}\exp(\boldsymbol{p}_i^{m}\cdot\boldsymbol{p}_i^{l}/\tau)},
% \end{equation}

\begin{equation}
    \label{eq:structure_aware_loss}
    L^{SA}=-\sum_{i=1}^{N}\sum_{m=1}^{2M}\log\frac{\exp(\boldsymbol{p}_i^{m}\cdot\boldsymbol{p}_{i}^{v(m)}/\tau)}{\sum_{l\neq m}\exp(\boldsymbol{p}_i^{m}\cdot\boldsymbol{p}_i^{l}/\tau)}
\end{equation} 
If image $m$ is an original image, then $v(m)$ represents its augmented view; if image $m$ is an augmented image, then $v(m)$ represents the original image.
 Minimizing $L^{SA}$ drives the representation pairs from the augmented images in the numerator close while pushing the representations in the denominator far away, where the denominator contains M-mode images from the same patient. 

Finally, we combine the two losses to get structure-aware and patient-aware contrastive loss for M-mode images using the hyperparameter $\alpha$ to control the trade-off between the awareness terms:
\begin{equation}
\label{eq:SSL_loss}
    L^{CL}=\alpha L^{PA} + (1-\alpha)L^{SA}.
\end{equation}

\section{Experiments and Results}
\subsection{Dataset}
We use the publicly available EchoNet-Dynamic dataset \citep{ouyang2019}. 
It contains $10'030$ apical-4-chamber echocardiography videos from individuals who underwent imaging between 2016-2018 as part of routine clinical care at Stanford University Hospital.
Each B-mode video was cropped and masked to remove information outside the scanning sector and downsampled into standardized $112 \times 112$ pixel videos. 
For simplicity, we used videos with at least 112 frames.
We use the official splits with $7465$ training, $1289$ validation, and $1282$ test set samples.

\subsection{Experimental Setup}
We evaluate the models' performance using classification accuracy for five random seeds and report the mean performance and standard deviation.
During training, all supervised models optimize the estimation of EF as a regression task.
For testing, we use a constant threshold $\tau$ for classifying cardiomyopathy. 
In all experiments, we set $\tau = 0.5$.
Hence, an estimation of $\hat{\tau} < 0.5 $ results in classifying a sample as cardiomyopathic.

We evaluate all models using the area under the receiver operating characteristic (AUROC) and the area under the precision-recall curve (AUPRC) with respect to whether a patient is correctly classified as healthy or cardiomyopathic.
Additionally, we report the mean absolute error (MAE) and the root mean squared error (RMSE) of the predicted EF with respect to the true EF in the Supplementary Material.
We report the mean performance, including standard deviations over five random seeds for all results.

We use the training set from EchoNet for pre-training (SSL), and apply a linear learning rate scheduler during the first 30 epochs as warm-up. 
For the supervised fine-tuning, we select different proportions of the training set in the limited labeled data scenario. 
All M-mode models are trained for $100$ epochs using Adam optimizer \citep{kingma2014adam} with an initial learning rate of $0.001$ and a batch size of $64$. 
For image augmentation, we apply random horizontal flip and Gaussian noise.
For the fusion method of the the M-mode representations we used concatenation.
For the EchoNet model, we use the same model and parameters as in \cite{ouyang2020}. 
The model is trained for $45$ epochs with a learning rate of $0.0001$ and a batch size of $20$. 
We do not use test-time augmentation for any of the models.
We report the full set of hyperparameters used in our experiments in \Cref{tab:hyperparameters}. 

\begin{table}
\caption{List the hyperparameters used in our experiments. We use the same hyper-parameters for E2E setup and the fine-tuning stage of SSL setup (denoted as "\_sup" in Table \ref{tab:hyperparameters}). "\_cl" denotes the hyper-parameters used in the SSL pre-training stage.}
\label{tab:hyperparameters}
    \centering
    \begin{tabular}{c|c|c}
        \toprule
        Parameter & Value & Description \\
        \midrule
        lr\_sup & $0.001$ & learning rate for supervised training \\
        lr\_cl & $1.0$ & learning rate for SSL training \\
        opt & Adam & optimizer for SSL and supervised training \\
        bsz\_sup & $64$ & batch size for supervised training \\
        bsz\_cl & $256$ & batch size for SSL training \\
        epoch\_sup & $100$ & epochs for supervised training \\
        epoch\_cl & $300$ & epochs for SSL training\\
        epoch\_warm & $30$ & warm-up epochs for SSL training \\
        $\alpha$ & $0.8$ & loss trade-off \\
        $\tau$ & $0.01$ & temperature scaling \\
        Dim\_e & $512$ & $Enc(\cdot)$ output dimension \\
        Dim\_ph & $2048$ & $Proj(\cdot)$ hidden layer dimension \\
        Dim\_po & $128$ & $Proj(\cdot)$ output dimension \\
        Dim\_lstm & $256$ & LSTM output dimension \\
        \bottomrule
    \end{tabular}
\end{table}

\subsection{Results and Discussion}
\label{sec:results_discussion}
\subsubsection{Evaluating M-mode Images in Supervised Setting} 
\begin{figure}
    \centering
    \begin{subfigure}[t]{0.9\textwidth}
        \includegraphics[width=\textwidth]{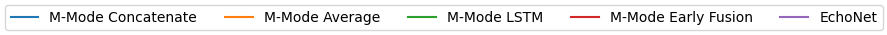}
        % \phantomcaption 
    \end{subfigure}
    \begin{subfigure}[t]{0.45\textwidth}
         \centering
         \includegraphics[width=\textwidth]{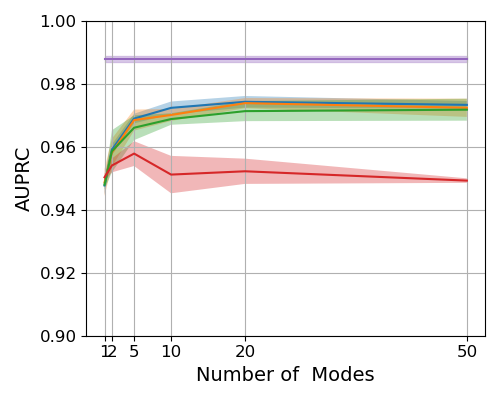}
         \caption{AUPRC}
         \label{fig:exp_mmode_num_modes_auprc}
     \end{subfigure}
     % \hspace{0.5cm}
     % \hfill
     \begin{subfigure}[t]{0.45\textwidth}
         \centering
         \includegraphics[width=\textwidth]{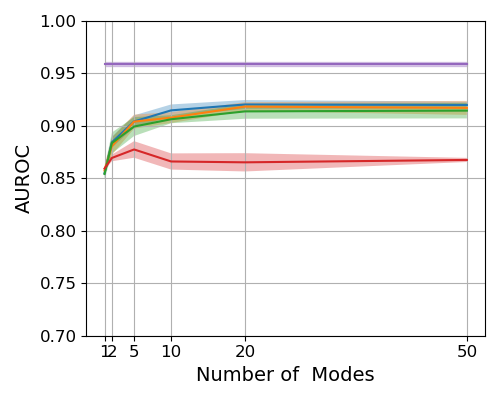}
         \caption{AUROC}
         \label{fig:exp_mmode_num_modes_auroc}
     \end{subfigure}
     \begin{subfigure}{0.45\textwidth}
         \centering
         \includegraphics[width=\textwidth]{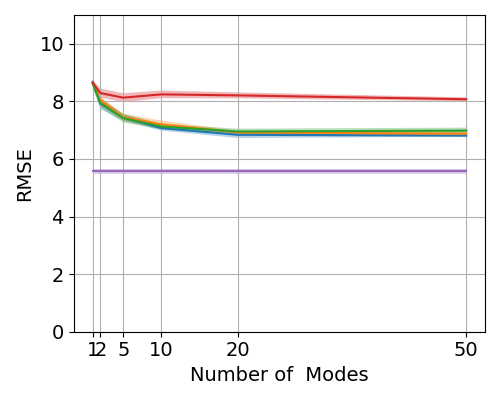}
         \caption{RMSE}
         \label{fig:exp_mmode_num_modes_rmse}
     \end{subfigure}
     \begin{subfigure}{0.45\textwidth}
         \centering
         \includegraphics[width=\textwidth]{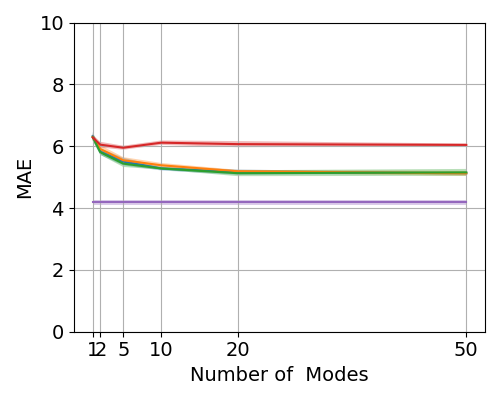}
         \caption{MAE}
         \label{fig:exp_mmode_num_modes_mae}
     \end{subfigure}
     \begin{subfigure}{0.45\textwidth}
         \centering
         \includegraphics[width=\textwidth]{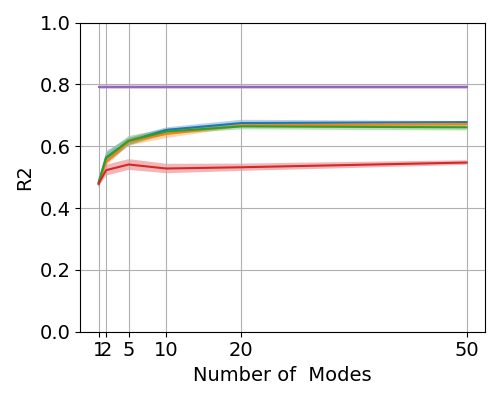}
         \caption{$R^2$}
         \label{fig:exp_mmode_num_modes_r2}
     \end{subfigure}
    % \hfill
     \caption{Performance for different numbers of M-mode images using early and late-fusion methods. In (a), we evaluate the classification performance with respect to AUPRC and AUROC in (b), the regression performance with respect to RMSE in (c), MAE in (d), and $R^2$-score in (e). 
     }
    \label{fig:exp_mmode_e2e_num_modes_classification}%
\end{figure}

We train and evaluate models with different numbers of M-modes for $M \in \{ 1, 2, 5, 10, 20, 50 \}$.
We use the complete training set, including labels, as we are interested in the performance of the models depending on the number of available M-modes.
\Cref{fig:exp_mmode_e2e_num_modes_classification} shows the results for different numbers of M-modes.
We see that late fusion models benefit from an increasing number of modes,
whereas the early fusion method overfits quickly and never achieves a comparable performance.

\subsubsection{Evaluating Limited Data Regime} 
We evaluate the accuracy of the different models introduced in \Cref{sec:methods} for different amount of labeled training samples.
As most medical datasets do not have the size of EchoNet-Dynamic \citep{kiryati2021}, methods for medical machine learning should perform best in the limited labeled data regime.
We use \emph{E2E} for the supervised and \emph{CL} for the self-supervised setting. 

Additionally, we introduce \emph{E2E+} and \emph{CL+}, which, inspired by EchoNet \cite{ouyang2020}, uses random short clips for each training epoch. 
Both models use M-mode images of 32 frames with a sampling period of 2. 
We train and evaluate models using $p\%$ of the full training set for $p \in \{ 1, 2, 3, 5, 10, 20, 30,$ $50,75, 100 \}$.
All M-mode methods are trained with $M=10$.

\begin{figure}[t]
    \centering
    \begin{subfigure}[t]{0.82\textwidth}
        \includegraphics[width=\textwidth]{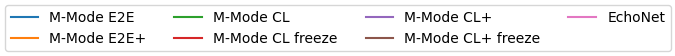}
        % \phantomcaption 
    \end{subfigure}
    \hfill
    \begin{subfigure}{0.45\textwidth}
         \centering
         \includegraphics[width=\textwidth]{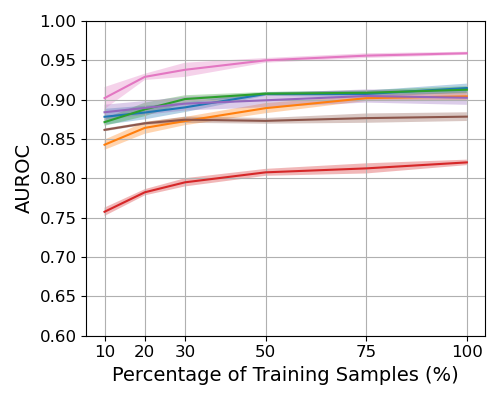}2
         \caption{10\% - 100\%}
         \label{fig:all_exp_limited_data_auroc_10_to_100}
     \end{subfigure}
     \begin{subfigure}{0.45\textwidth}
         \centering
         \includegraphics[width=\textwidth]{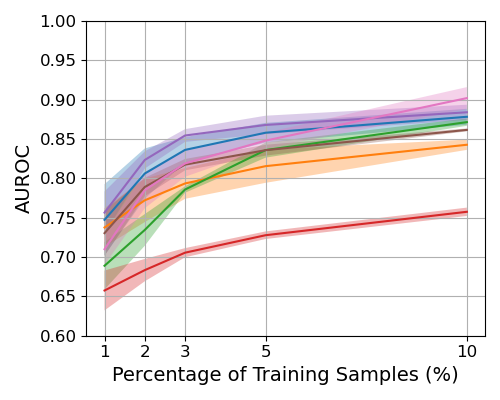}
         \caption{1\% - 10\%}
         \label{fig:app_exp_limited_data_auroc_1_to_10}
     \end{subfigure}
    % \hfill
     \caption{
     Results for different training set sizes using the proposed end-to-end supervised (E2E) and contrastive learning (CL) approaches.
     In (a), we train and evaluate the models on 10\%-100\% labeled training samples,
     in (b) only on 1\%-10\% of the samples.
     E2E and CL models are trained using a fixed long clip with length 112; E2E+ and CL+ are trained using random short clips with length 32. CL freeze and CL+ freeze are fine-tuned with the encoder parameters frozen.
     }
    \label{fig:app_exp_limited_data_auroc}%
\end{figure}

\Cref{fig:app_exp_limited_data_auroc} shows the limited labeled data experiment results.
Although we are not able to reach the performance of the EchoNet model for any number of modes (see \Cref{fig:exp_mmode_num_modes_auroc}) if the number of labeled training samples is high (see \Cref{fig:all_exp_limited_data_auroc_10_to_100}), both supervised and self-supervised learning methods using M-mode instead of B-mode can outperform the EchoNet model in the low labeled data regime ($p < 5\%$, \Cref{fig:app_exp_limited_data_auroc_1_to_10}). 
Also, we observe that using shorter clips is useful for the self-supervised learning methods, with \emph{CL+} being able to achieve an AUROC over $0.85$ with only around $200$ labeled samples.

\subsubsection{Computational Cost} 
Furthermore, we compare the number of parameters and computational costs for different models in \Cref{tab:cost}, where we used a multi-GPU setup with four NVIDIA GeForce RTX 2080 Ti GPUs. 
We report the computation time in seconds per batch (sec/B) and milliseconds per sample (msec/sample), and the memory requirements in gigabytes per batch (GB/B).

Our proposed M-mode image based models require around six times less time and ten times less memory to train and run inference per sample. 
Given the used memory per batch, we could increase the batch size for the M-mode methods, lowering the computation time per sample even further, whereas the baseline model is already at the limit due to its architecture. 

\begin{table}[b]
\caption{Computational costs.
We evaluate the EchoNet and the proposed M-mode methods with respect to the number of parameters, the computation time, and the memory requirements.
All M-mode models are evaluated using $M=10$.
E2E defines the end-to-end supervised and CL the contrastive learning approach.
}
\label{tab:cost}
    \centering
    \begin{tabular}{lcccccccc}
        \toprule
         & & & \multicolumn{2}{c}{Time (sec/B)} & \multicolumn{2}{c}{Time (msec/sample)} & \multicolumn{2}{c}{Memory (GB/B)}  \\
         \cmidrule(r){4-5} \cmidrule(r){6-7} \cmidrule(r){8-9}
         Model & BS & \#Params (Mio.) & Train & Test & Train & Test & Train & Test \\
        \midrule
        EchoNet & 20 & 31.5 & 2.898 & 2.474 & 144.9 & 123.7 & 5.294 & 1.187 \\
        E2E \& CL & 64 & 11.7 & 1.568 & 1.330 & 24.5 & 21.1 & 1.013 & 0.120 \\
        % CL-freeze (bsz=64) & 11.7 & 1.395 & 1.330 & 0.177 & 0.120 \\
        \bottomrule
    \end{tabular}
\end{table}

\section{Discussion and Conclusion}
In this work, we propose to generate M-mode images from readily available B-mode echocardiography videos and fuse these to estimate EF and, thus, cardiac dysfunction. 
Our results show that M-mode-based prediction methods are comparable to the baseline method while avoiding its complex training routine and reducing the computational cost and the need for expensive expert input.

Conventional M-mode images have a very high sampling rate, which results in a high temporal resolution so that even very rapid motion can be recorded. 
The generated M-mode images have significantly less temporal resolution than the conventional M-mode images from US machines. 
However, our results indicate that exploiting generated M-mode images does not limit the performance for EF estimation. 
As we do not use the M-mode images collected directly from the US machines, there is no need for an additional data collection step. 

Additionally, we show the potential of pre-trained methods.
In scenarios where expensive expert labels are not readily available, pre-training using unlabeled M-mode images outperforms more complicated pipelines highlighting the potential of M-Mode based pipelines for clinical use cases.
In our future work, we want to investigate the use cases for M-mode on different diseases and further improve the performance of the proposed pre-training pipeline. 

\section{Acknowledgements}
EO was supported by the SNSF grant P500PT-206746 and 
TS by the grant 2021-911 of the Strategic Focal Area “Personalized Health and Related Technologies (PHRT)” of the ETH Domain (Swiss Federal Institutes of Technology).

\bibliographystyle{splncs04}
\bibliography{053-main.bib}

\end{document}